\documentclass[aps,prl,twocolumn,showpacs,superscriptaddress]{revtex4}
\usepackage{graphicx}

\begin{document}

\title{Complete \textit{d}-Band Dispersion and the Mobile Fermion Scale in Na$_x$CoO$_2$}
\author{D. Qian}
\author{L. Wray}
\author{D. Hsieh}
\affiliation{Department of Physics, Joseph Henry Laboratories,
Princeton University, Princeton, NJ 08544}
\author{L. Viciu}
\affiliation{Department of Chemistry, Princeton University,
Princeton, NJ 08544}
\author{R.J. Cava}
\affiliation{Department of Chemistry, Princeton University,
Princeton, NJ 08544}
\author{J.L. Luo}
\author{D. Wu}
\author{N.L. Wang}
\affiliation{Institute of Physics, Chinese Academy of Sciences,
Beijing 100080, China}
\author{M.Z. Hasan}
\affiliation{Department of Physics, Joseph Henry Laboratories,
Princeton University, Princeton, NJ 08544}

\begin{abstract}

We utilize fine-tuned polarization selection coupled with
excitation-energy variation of photoelectron signal to image the
\textit{complete d}-band dispersion relation in sodium cobaltates. A
hybridization gap anticrossing is observed along the Brillouin zone
corner and the full quasiparticle band is found to emerge as a
many-body entity lacking a pure orbital polarization. At low
dopings, the quasiparticle bandwidth (Fermion scale, many-body
\textit{E$_F$} $\sim$ 0.25 eV) is found to be smaller than most
known oxide metals. The low-lying density of states is found to be
in agreement with bulk-sensitive thermodynamic measurements for
nonmagnetic dopings where the 2D Luttinger theorem is also observed
to be satisfied.

\end{abstract}


\date{ Submitted to Phys.Rev.Lett. on June-28th, 2006}

\maketitle

Sodium cobaltate (Na$_{1/3}$CoO$_2\cdot y$H$_2$O) is the only known
example of an oxide superconductor, besides high Tc cuprates, which
is realized in the vicinity of a spin-1/2 Mott state. However,
unlike the single band cuprates, the layered cobaltate class
Na$_x$CoO$_2$ is a prime example of a multiorbital interacting
electron system. Observations of superconductivity, magnetic order,
spin-thermopower, charge-order, non-Fermi-liquid transport, and many
other unusual properties\cite{1,2,3} have led to an extensive body
of theories to model the electron
behavior\cite{4,5,6,7,8,9,10,11,12,13,14,15,16,17,18}. The
symmetries proposed for the superconducting order
parameter\cite{4,5,6,7} critically depend on the orbital character
of the bands and the Fermi surface topology. Moreover, a measure of
the quasiparticle bandwidth is crucial for the theories of enhanced
thermopower and anomalous Hall effect\cite{6,8}.

Previous angle-resolved photoemission spectroscopy (ARPES)
measurements\cite{9,10,11} have focused on the Fermi surface
topology, Luttinger theorem, and low-energy physics. However, many
important issues such as the orbital polarization of the
quasiparticles, the reason for the sinking of the e$_g'$ bands, and
the effects of \textit{hydration} have not yet been explored. The
experimental map of the Fermi surface (FS)\cite{9,10,11} of these
compounds revealed only one FS. The band associated with the corner
pockets predicted in local-density approximation (LDA)\cite{12} was
observed to have sunk\cite{10,11} below the Fermi level. Three
different intrinsic many-body explanations have been advanced to
account for this unusual phenomenon. One scenario argues that the
corner pockets sink due to a delicate balance of interorbital
\textit{Coulomb correlations}\cite{16,17}. A second
scenario\cite{18} suggests that the effect of the random potential
of the sodium layer leads to a weak localization of the pocket
states. Finally, it has been suggested that the \textit{polarization
employed} in all previous ARPES studies precludes the detection of
the pockets [15]. In this Letter, we report high-resolution
measurements of the \textit{complete d}-band multiplet, which
enables a comparison of the data with theoretical models in the
literature.

\begin{figure}[b]
\center \includegraphics[width=9cm]{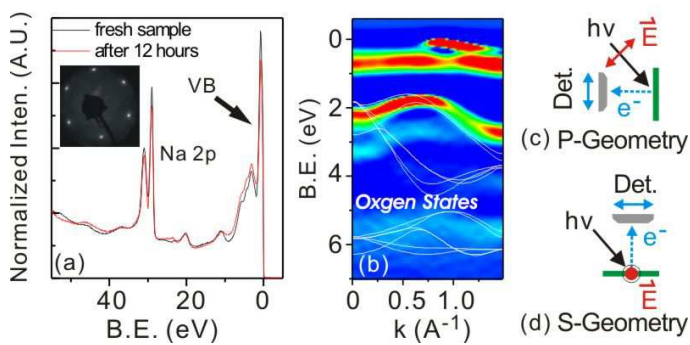} \caption{Sample
characterization: (a) Na-2p core level signal is monitored during
the experiment. Inset shows the LEED image exhibiting a clear
sixfold symmetry of the in situ cleaved surface. No ruthenatelike
surface reconstruction is observed. (b) No significant shift of the
oxygen states are observed at the surface. Experimental
configurations employed: Polarization of light is parallel (c) or
perpendicular (d) to the plane defined by the sample and the
orientation of the detector slits.}
\end{figure}

\begin{figure*}[t]
\center \includegraphics[width=16cm]{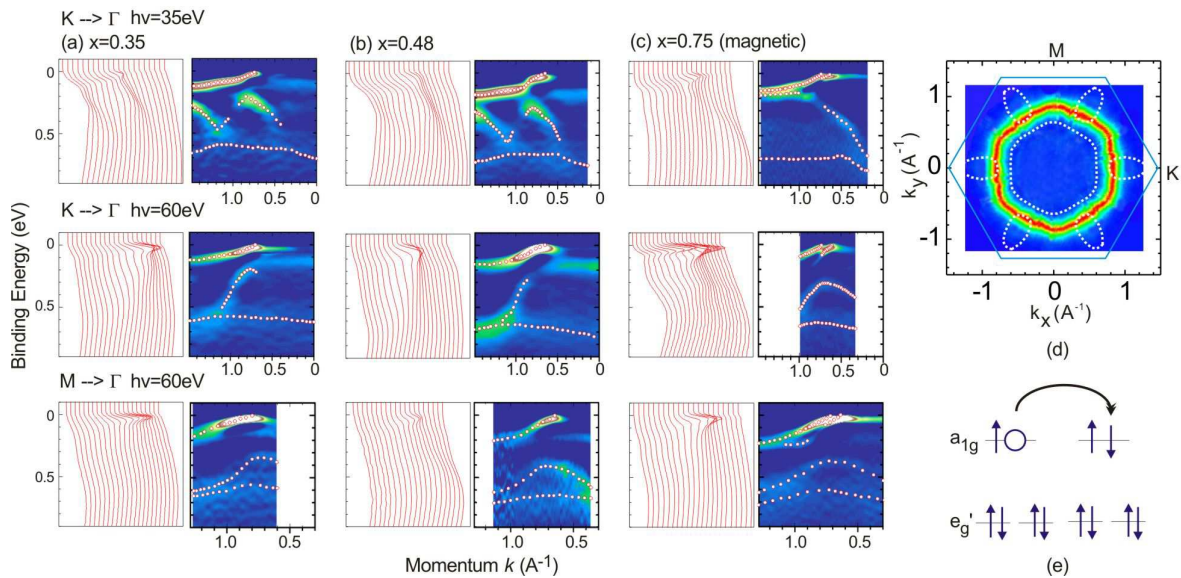} \caption{Doping
evolution of \textit{d}-band complex: Electron band structures of
$t_{2g}$ complex measured along $\Gamma$-K (top and middle rows) and
$\Gamma$-M (bottom row) for doping levels of (a) x=0.35, (b) x=0.48,
and (c) x=0.75. Each panel includes energy distribution curves
(left) and corresponding secondary derivative image plot (right).
The bands are marked using red dots. Bands along $\Gamma$-K exhibit
strong photon energy dependence. Data for two photon energies are
shown: $hv$ = 35 eV (top) and $hv$ = 60 eV (middle). No significant
photon energy dependence of bands were found along the $\Gamma$-M
cut. The extended flat feature (unmarked) past the Fermi crossing in
image plots for low dopings are due to rapid drop of background. (d)
A high-resolution Fermi surface shows the k-space cuts along
$\Gamma$-K and $\Gamma$-M. (e) A cartoon view of hole doping.}
\end{figure*}

We have extended the electronic structure determination problem in
three regards: First, in order to accurately image the band
dispersion over the complete Brillouin zone (BZ) we utilized a range
of incident photon energies. Second, we employed two distinct pure
polarization settings, namely, the \textit{S} and the \textit{P}
scattering geometries. Finally, being equipped with these methods,
we studied the effect of hydration on the electronic structure.
Spectroscopic measurements were performed with 30 to 90 eV photons
with better than 12 to 25 meV energy resolution, and angular
resolution better than 0.5\% of the Brillouin zone at ALS beam lines
10.0.1 and 12.0.1 using Scienta analyzers with chamber pressures
better than 4$\times$10$^{-11}$ torr. A rotatable Scienta analyzer
allowed us to work under the \textit{P}-geometry [Fig. 1(c)] or
\textit{S}-geometry [Fig. 1(d)] polarization mode or under a mixed
but tunable configuration. Carefully grown (by two different
methods, flux and traveling solvent floating zone) high-quality
single crystals over a wide doping range x=0.3, 0.35, 0.48, 0.57,
0.7, and 0.75 were used for this study. Cleaving the unhydrated
samples \textit{in situ} at 20 K (or 100 K) resulted in shiny flat
surfaces, characterized by electron diffraction to be clean and well
ordered with the same symmetry as the bulk [Fig. 1(a)]. The x=0.35
doping was further studied in hydrated (Na$_{1/3}$CoO$_2\cdot
y$H$_2$O) and unhydrated (Na$_{1/3}$CoO$_2$) forms. The hydrated
sample surface was prepared by cleaving at 15 K (much below the
freezing temperature of H$_2$O) for the frozen-in retention of
hydration. No pressure bursts were observed upon cleaving. Moreover,
samples were covered with silver paste to reduce lateral outgassing
of H$_2$O.

\begin{figure}[b]
\center \includegraphics[width=9cm]{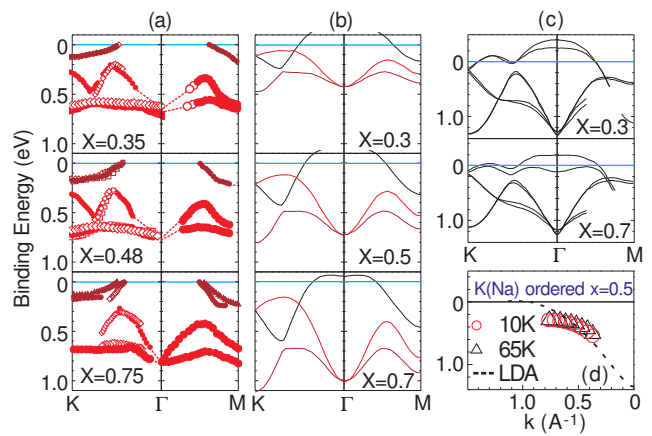} \caption{
Comparison of (a) experimental band structures for x =0.35, x=0.48,
and x=0.75 with (b) LDA + large-\textit{U} calculation\cite{17} and
(c) LAPW calculations\cite{14}. (d) For sodium- or potassium-ordered
samples, x=1/2, the $e_g'$ band is always below the Fermi level
($\sim$250 meV) at the temperature below or above insulating
transition points.}
\end{figure}

Figure 2 shows the doping dependence of band dispersions of the
$t_{2g}$ complex measured as a function of incident photon energy.
Although a quasiparticle is observed to cross the Fermi level
independent of the photon energy or crystal direction used for the
measurement, the deeper-lying bands resonate at different photon
energies. This sensitivity is remarkable for the high-energy bands,
especially along the $\Gamma$-K cut. At some photon energies, part
of the bands are apparently "missing" or exhibit very small cross
section. Therefore to image the full band structure we have explored
the bands at many incident photon energies between 30 to 90 eV. The
representative data are shown in Fig. 2. Band dispersions are
plotted in Fig. 3(a). We note that along all cuts the quasiparticle
band that crosses the Fermi level is well separated from the rest.
Along the $\Gamma$-M cut, there are essentially two branches of the
deeper-laying bands. These bands can be traced back to their LDA
origin and be labeled as the $e_g'$ bands\cite{12}. With increasing
doping they sink below the Fermi level in a systematic way and never
cross each other except meeting or emanating from the $\Gamma$
point. Along the $\Gamma$-K cut, a nearly flat band is observed
around 0.6 eV. Then in between there is an undulatory band with peak
near $k \sim$ 0.55$\pi$. The absence of dramatic photon energy
dependence for k$_z$ dispersion of the double band behavior observed
in high doping (x=0.75 or higher doping\cite{11}) suggests the lack
of three-dimensional coupling and points to a ferromagnetic or phase
separated sample origin of the observed behavior.

We now compare our data with the calculations in\cite{17} (Fig. 3).
First, we note that all deeper-lying bands meet at the $\Gamma$
point, consistent with theory and as expected on fundamental
$t_{2g}$ symmetry considerations. Second, a reasonable agreement is
seen along the $\Gamma$-M cut in terms of the connectivity of the
deeper-lying bands. Along the $\Gamma$-K cut, however, theory
predicts a crossing between the upper $e_g'$ band ($e_{g1}$) and the
$a_{1g}$ quasiparticle band. Our data suggest that this crossing is
avoided at all dopings, and indicate a gap between the $a_{1g}$ band
and the upper $e_g'$ band near $k \sim 0.55\pi$. Previously, this
effect showed up only as a break in dispersion and was interpreted
as a "kink" due to electron-phonon coupling\cite{9,10}. In our
high-quality data set it appears, instead, to resolve as a multiband
hybridization effect [Fig. 3(a) (left)]. Our data suggest a strong
mixing of the bands leading to an \textit{anticrossing} behavior and
the opening of a hybridization gap of magnitude more than 100 meV
[Fig. 3(a) (left)]. This also suggests that the full quasiparticle
band cannot be of pure $a_{1g}$ symmetry.

\begin{figure}[b]
\center \includegraphics[width=7cm]{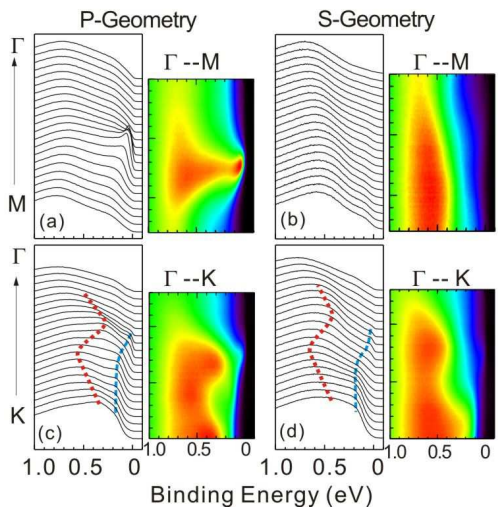}
\caption{Polarization dependence: The $t_{2g}$ complex measured
along the $\Gamma$-M and $\Gamma$-K cuts under \textit{P} geometry
(a),(c) and \textit{S} geometry (b),(d). Red and blue (light and
gray) dots trace the $e_g'$ and $a_{1g}$ bands, respectively.
Geometries are defined in Figs. 1(c) and 1(d).}
\end{figure}

We now consider the selection rules associated with $e_g'$ and
$a_{1g}$ bands for different ARPES geometries. A symmetry analysis
suggests that the $e_{g1}$ states (based on the basis functions in
LDA or DMFT\cite{15}) near $E_F$ and along the $\Gamma$-K cut are
odd with respect to reflections about the plane formed by $\Gamma$-K
and the normal of the Co plane. These $e_{g1}$ states at $E_F$ along
$\Gamma$-K cannot possibly be observed using \textit{P}-polarized
geometry with coinciding planes of incidence and emission. The even
$a_{1g}$ and $e_{g2}$ bands (the lower $e_g'$ band) exhibit
hybridization gaps. The odd $e_{g1}$ band crosses these bands
($a_{1g}$ and $e_{g2}$) without any interaction. For $x$ along
$\Gamma$-K and y along $\Gamma$-M, emission in the $xz$ plane
containing $\Gamma$K and \textit{P}-polarized light incident in the
same plane selection rules precludes any contribution from $e_{g1}$
states since the final state is even in $y$. The detection geometry
for the photoelectrons permits excitations only from $a_{1g}$ and
$e_{g2}$ bands. Emission from $e_{g1}$ states is allowed for the
\textit{S}-polarized light incident in the $xz$ plane or the
\textit{P}-polarized light incident in the $yz$ plane. Figures 4(a)
and 4(b) show the experimental data along the $\Gamma$-M cut using
\textit{P} and \textit{S} geometries, respectively. Only with the
\textit{P} geometry is strong emission detected near the $\Gamma$-M
crossings. This is consistent with an $a_{1g}$ character of the
low-lying states along this cut. Figures 4(c) and 4(d) show the
$\Gamma$-K cut using \textit{P} and \textit{S} geometries,
respectively. The emission contrast is much less dramatic and the
quasiparticle is only weakly enhanced under the \textit{P} geometry.

\begin{figure}[b]
\center \includegraphics[width=8cm]{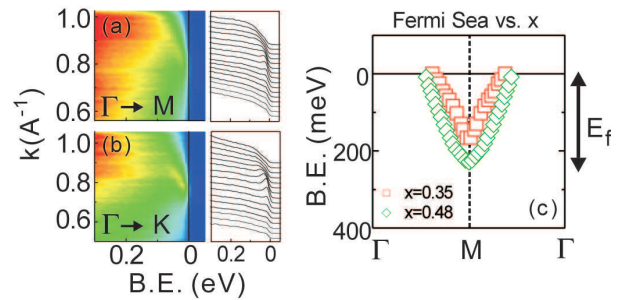} \caption{Effect of
hydration near x=1/3: (a),(b) Fermi crossings are observed at 15 K
in the hydrated Na$_x$CoO$_2$.yH$_2$O compound near x=0.35 while
"$y$" varies from 0.7 to 1.4 (see text). (c) Quasiparticle bandwidth
(many body $E_F$) is estimated from a complete dispersion relation
measurement.}
\end{figure}

We note that LDA plus the multiorbital mean-field Hubbard model (LDA
+ large-\textit{U}) can properly account for the sinking of the
$e_g'$ pockets and the suppression of the bandwidth at low Na doping
near $x = 0.3$\cite{17}. Experiment suggests that a hybridization
gap is also involved consistent with the suggestion by Singh
\textit{et al.}\cite{18}.

Another mechanism that has been proposed for the sinking of the
$e_g'$ states is the effect of the random potential from the Na
layer. This model suggests that the disordered sodium potential
experienced by the heavy $e_g'$ states leads to a localization of
the carriers. We have studied the high-energy band structure of
(Na/K)$_{1/2}$CoO$_2$ samples where Na or K ions order well above
room temperature, thus removing the random potential which is
experienced by the $e_g'$ electrons at all other dopings.
Experimentally, we find that while $a_{1g}$ bands exhibit a gap of
size less than 10 meV only at temperatures below the metal insulator
transition ($\sim$ 51 K), the $e_g'$ bands exhibit a clear gap well
above the transition temperature and are located at about 250 meV
below $E_F$ [Fig. 3(d)] similar to what is seen at other dopings
where Na ions are highly disordered. Therefore our results suggest
that the sinking of the $e_g'$ bands is not related to weak
localization only. We further compare our data with linear
augmented-plane-wave method (LAPW) calculations\cite{14} in Fig.
3(c). LAPW results agree with our observation of a hybridization gap
along $\Gamma$-K.

We have studied the effect of hydration on the low-lying electronic
structure. Results are shown in Fig. 5. In comparison with
unhydrated samples, our results suggest that hydration does not lead
to any significant change in Fermi velocity. Under an unfavorable
scenario where some H$_2$O is evaporated from top surface (unlikely
to happen since experiments are carried out well below the freezing
temperature of H$_2$O) a partial hydration effect is still probed in
ARPES due to the buried $n$H$_2$O layers.

Our detailed systematic high-resolution measurements in the current
study allow us to draw a direct connection with bulk-sensitive
electronic density of states measurements. We note the average size
of the Fermi surface k$_F$ $\sim$ 0.77 $\pm$ 0.05 \AA$^{-1}$ and the
velocity $\hbar$v$_F$ $\sim$ 0.4 $\pm$ 0.1 eV$\cdot$\AA \ for $x$
$\sim$ 1/3. Since the samples are mostly two dimensional in this
doping range, the linear specific heat coefficient, $\bf{\gamma}$,
is set by $k_F$ and $v_F$ only as shown in \cite{19,20,21}:
$\bf{\gamma}$ = [($\pi$N$_A$k$_B^2$a$_o^2$)/(3$\hbar$$^2$)]$\sum$m*,
where m* = $\hbar$k$_F$/v$_F$ (N$_A$ is the Avogadro's number, k$_B$
is the Boltzman's constant, $\hbar$ is the Planck's constant and
a$_o$ is the lattice constant) and $\bf{\gamma}$ $\sim$ 11.47 $\pm$
2.96 mJ/mol K$^2$. (a$_o$ = 2.82 $\AA$)

This ARPES estimated value is close to that obtained with bulk
thermodynamic measurements ($\sim$ 12.1 for unhydrated and 13.5
mJ/mol K$^2$ for the hydrated or the deuterated superconductors
\cite{22}). This leaves very little room for the six corner pockets
to be shared, and if they exist, their sizes must be much smaller
than the best k resolution of ARPES. Therefore, our data suggest
that almost all of specific heat can be consistently explained with
a single Fermi surface with a narrow renormalized band as we observe
in this high-resolution work. This point is further supported by
Shubnikov-de Haas data\cite{19}. We also note that for x=0.75 or
higher dopings\cite{11}, ARPES data and specific heat data differ by
more than a factor of 2. This discrepancy could be due to magnetic
order on the top layer and/or intrinsic phase separation recently
reported by several groups in these magnetic dopings\cite{23}. This
is the regime where 2D nonmagnetic Luttinger theorem was found to be
not satisfied\cite{11}.

Finally, our systematic results allow us to estimate the bandwidth
of the band that generates the FS in a clear fashion as shown in
Fig. 5(e) since for the first time it can be traced over all of the
BZ. The effective bandwidth of the occupied band that crosses the
Fermi level is on the order of 0.25 eV. This is smaller than most
other correlated oxide metals such as the cuprates, ruthenates or
manganites\cite{23,24}.

In conclusion, we have measured the complete band dispersions in
sodium cobaltates by utilizing the polarization and
excitation-energy dependent matrix elements. A hybridization gap is
observed along $\Gamma$-K which was previously interpreted as strong
electron-phonon coupling. The extracted effective bandwidth, based
on our complete dispersion measurements, is found to be smaller than
most other similar correlated oxides. The measured ARPES density of
states is found to be in agreement with the bulk electronic density
of states measured in the specific heat and consistent with a single
Fermi surface in the doping range where the 2D Luttinger theorem is
fully satisfied in the ARPES data.

We acknowledge G. Khaliullin, G. Kotliar, B. Keimer, P. A. Lee, N.
P. Ong, D. Singh, Z.Wang, and A. Fedorov for discussions. This work
is primarily supported by DOE Grant No. DE-FG-02-05ER46200. R. J.C.
acknowledges support by the NSF (DMR-0213706). N. L.W. acknowledges
support by the NSFC (10574158).

\end{document}